\documentclass[]{article} 
\usepackage{amsmath}
\usepackage{graphicx}
\usepackage[top=2.cm, bottom=2.cm, left=2.cm, right=2.cm]{geometry}

\begin{document}
\title{Analysis of Two-Particle Systems in $2+1$ Gravity Through Hamiltonian Dynamics}
\author{Alexandre Yale\footnote{ayale@perimeterinstitute.ca} \footnote{University of Waterloo, Waterloo, Ontario, Canada} \footnote{Perimeter Institute, Waterloo, Ontario, Canada}~,  
				R. B. Mann\footnotemark[2] \footnotemark[3]~ and 
				Tadayuki Ohta\footnote{Miyagi University of Education, Sendai, Miyagi, Japan}} 
\date{November 25, 2010}

\maketitle
\newcommand{\vecvar}[1]{\mbox{\boldmath$#1$}}

\abstract{We study the dynamics of particles coupled to gravity in $(2+1)$ dimensions.  Using the ADM formalism, we derive the general Hamiltonian for an $N$-body system and analyze the dynamics of a two-particle system.  Nonlinear terms are found up to second order in $\kappa$ in the general case, and to every order in the quasi-static limit. }

\section{Introduction }

The study of 2+1 dimensional gravity \cite {Menotti, Staruszkiewicz, Leutwyler, Deser, Gott, Giddings, Clement, Clement2, Grignani} has been around for several decades.  Its attraction was rooted in its mathematical simplicity, which afforded some insight into the dynamics and behaviour of general relativistic gravity.   Indeed, in this framework, the Einstein tensor can be expressed in terms of the curvature tensor, such that vacuum must be locally flat.  This framework became even more popular after the discovery of the BTZ black hole \cite{Banados, Banados2} and is now a standard tool practitioners of quantum gravity employ in understanding their subject
 \cite{Banados3, Carlip, Carlip2}.

We are concerned here with the problem of $N$-body dynamics, which is a long-standing one in relativity due to its notorious difficulty.  In  lower dimensional settings, this problem becomes much simpler. For example, the general form of the solution has been obtained for lineal gravity \cite{Ohta} for arbitrary $N$, after which a variety of exact solutions for $N=2$ were obtained in various contexts that include both charge and cosmological expansion/contraction \cite{Mann, Mann2, Mann3, Mann5}, by investigating the Hamiltonian of such a system through canonical reduction.   Several interesting exact solutions to the $N$-body equilibrium problem \cite{Mann4,Mann6,Ryan1,Ryan2} in (1+1) dimensions have also been obtained.

In this paper, we follow a similar approach in (2+1) gravity to obtain canonical equations of motion to analyze a two-body system.  The analysis of the $N$-body problem in (2+1) dimensions also has an interesting history, beginning with construction of a spinning point-particle solution \cite{Deser} and then a consideration of the quantum scattering problem \cite{Sousa}.  Further developments came upon realizing that the problem could be analyzed from a topological perspective \cite{Bellini}, and an implicit solution for the metric and the motion of $N$ interacting particles was obtained \cite{Bellini2}.  Based on a mapping from multivalued Minkowskian coordinates to single-valued ones, it becomes explicit for two particles with any speed and for any number of particles with small speed. It is possible to show that the collision of point particles in 2+1 AdS spacetime can result in the formation of a black hole \cite{Matschull}.

The connection between these approaches and more traditional canonical methods  as employed in (1+1) \cite{Ohta} and (3+1) dimensions \cite{Ohta2} has not been explicated. In this paper, we address this issue. We begin by finding the total action corresponding to the system, which consists of the Einstein-Hilbert action for the field as well as a term corresponding to coupling gravity to matter.  A variational approach  will then lead to coordinate conditions and constraints, which will in turn produce the total Hamiltonian as an expansion in powers of the gravitational coupling $\kappa$.  The Hamiltonian will be explicitly calculated to second order in $\kappa$ for the general two-body case and to every order for the quasi-static limit.  Our results agree with previous work in (2+1) dimensions which has explored, through geodesics, the equations of motion in the quasi-static approximation \cite{Bellini}.

\section{Action}
\label{sec:action}
A first step is to derive the action for an $N$-body system.  This action will consist of two parts: the first, $I_E$, is the Einstein-Hilbert part due to the geometry of spacetime, and the second, $I_M$, is due to matter-gravity coupling.  Our derivation will follow the general ideas from the original ADM article \cite{ADM}.  In this formalism, the metric is defined as
\begin{equation}\label{admform}
	ds^2 = -N^2_0dt^2  + g_{ij}\left(dx^i + N^i dt\right)\left(dx^j + N^j dt\right)
\end{equation}
where $N_0 = \left(-g^{00} \right)^{-\frac{1}{2}}$ and $N_i = g_{0i}$ are the lapse function and shift covector.  We also define the extrinsic curvature $K_{ij}=(2N_0)^{-1} \left( N_{i|j}+ N_{j|i}-g_{ij,0} \right)$ and the canonical momentum conjugate to the metric: $\pi^{ij}=-\sqrt{g}(K^{ij}-Kg^{ij})$, where the vertical bar denotes a covariant derivative with respect to the induced metric $g_{ij}$.  Using the Gauss-Codazzi equations, we can perform a $2+1$ decomposition of the Einstein-Hilbert action by writing $^3R$ as a combination of $^2R \equiv R$ and extrinsic curvatures: %
\begin{equation} \begin{split}
	I_E&= \frac{1}{2\kappa} \int d^3x \sqrt{-^3g}{\ } ^3R \\
	&= \frac{1}{2\kappa} \int d^3x N_0 \sqrt{g} \left(R - K^2 + K_{ij}K^{ij}\right) - (2 \sqrt{g} K)_{,0} + \left[2 \sqrt{g} (K N^i - g^{ij}N_{0,j})\right]_{,i}\\
	&= \frac{1}{2\kappa} \int d^3x N_0 \sqrt{g} R - \frac{N_0}{\sqrt{g}}(\pi^2 - \pi_{ij} \pi^{ij}) - 2\pi_{,0}+2\left[\pi N^i - \sqrt{g} g^{ij}N_{0,j}\right]_{,i}\\
	&=\frac{1}{2\kappa}\int d^{3}x\left\{\pi^{ij}g_{ij,0}+N_0R^0+N_iR^i-2\pi_{,0}-2[\pi^{ij}N_{j}-\pi N^{i}+\sqrt{g}g^{ij}N_{0,j}]_{,i}\right\}.\\
	 &=\frac{1}{2\kappa}\int d^{3}x\left\{\pi^{ij}g_{ij,0}+N_{0}R^{0}+N_{i}R^{i}\right\},
\end{split} \end{equation}
where $R^0=\frac{1}{\sqrt{g}}(gR+\pi^2-\pi_{ij}\pi^{ij})$ and $R^i=2\pi^{ij}_{|j}$, and where in the last step we discarded total divergences.    We now calculate the Lagrangian corresponding to matter-gravity coupling, which will give us the second half of the action.  We consider the Lagrangian which minimally couples $N$ particles to gravity:
\begin{equation} \label{geod}
{\cal L}_M = 
\sum_{a}\int ds_{a}\left\{ -m_{a}\sqrt{\left( -g_{\mu \nu }(z)\frac{%
dz_{a}^{\mu }}{ds_{a}}\frac{dz_{a}^{\nu }}{ds_{a}}\right)}
\right\} .
\end{equation} 
We can rewrite this Lagrangian in a more convenient form by introducing the canonical momentum $p_{a \mu} = \frac{\partial {\cal L}_M}{\partial \dot{z}^\mu_a}$ conjugate to $z_{a \mu}$ as well as Lagrange multipliers $\lambda_a$ which ensure $p_a^\mu p_{a \mu} = -m_a^2$: 
\begin{equation} \begin{split}
	{\cal L}_M &=\sum_a \int ds_a \left(p_{a\mu} \frac{dz_a^\mu}{ds_a}-\frac{1}{2}\lambda_a'(s_a) (p_{a\mu}p_{a\nu}g^{\mu\nu}(x)+m^2_a)\right) 	\delta^{(3)}(\mathbf{x}-\mathbf{z_a}(s_a))\\
	&=\sum_a\left(p_{a\mu} \dot{z}^{\mu}_{a}-\frac{1}{2}\lambda_a (p_{a\mu}p_{a\nu}g^{\mu\nu}(x)+m^2_a)\right) 	\delta^{(2)}(\mathbf{r_a}(x^0))\\
&= \sum_a\left(p_{ai}\dot{z}^i_a -N_0 \sqrt{p_{ai}p_{aj}g^{ij}+m^2_a} +N_ip_{aj}g^{ij}\right) \delta^{(2)}(\mathbf{r_a}(x^0))
\end{split} \end{equation}
where $\lambda_a = \lambda'_a \frac{ds_a}{dz^0_a}$ (because of the $\delta$-function integration) and $\mathbf{r_a} = \mathbf{x}-\mathbf{z_a}$, and where in the last step we used (\ref{admform}) as well as the solution to the constraint $p_a^\mu p_{a \mu} = -m_a^2$: 
\begin{equation}
	p_{a0}=N_i p_{aj} g^{ij} - N_0 \sqrt{p_{ai}p_{aj}g^{ij}+m^2_a}.
\end{equation}
Combining this with our previous result for the Einstein-Hilbert action, we write the total action $I = I_E + \int {\cal L}_M$ as
\begin{equation}\label{ac-1}
	I = \int d^3x \left( \sum_a p_{ai} \dot{z}^i_a \delta^{(2)}(\mathbf{r_a}) + \frac{1}{2\kappa} ( \pi^{ij} g_{ij,0} + N_0 R^0 + N_i R^i ) \right),
\end{equation}
where \cite{Deser,ADM}
\begin{equation} \begin{split} \label{eqn:split}
	R^{0}&=\frac{1}{\sqrt{g}}(gR+\pi^2-\pi_{ij}\pi^{ij})-2\kappa\sum_a \sqrt{p_{ai}p_{aj}g^{ij}+m^2_a}\delta^{(2)}(\mathbf{r_a})\\
	R^{i}&=2\pi^{ij}_{|j}+2\kappa\sum_{a}p_{aj}g^{ij}\delta^{(2)}(\mathbf{r_a}) 
\end{split} \end{equation}
must both vanish following variations of $N_0$ and $N_i$.  These constraints (\ref{eqn:split}) have a very natural physical interpretation.  The latter of these is a momentum-balance constraint, indicating that in this  (2+1)-dimensional self-gravitating system the particle momenta must be cancelled by the momentum of the gravitational field.  The former is an energy-balance constraint, which basically indicates that the energy of the gravitational field (a combination of ``potential energy" expressed as the curvature of the spacelike slice and ``kinetic energy"  consisting of the square of the gravitational momentum)  must equal the relativistic energy of the particles.
\\\\
We now consider rewriting this action by choosing convenient coordinate conditions.  We define ${h}_{ij}=g_{ij} - \delta_{ij}$, where $\delta_{ij}$ is the Kronecker delta, and use the orthogonal decomposition 
\begin{equation} {h}_{ij}=\left(\delta_{ij}-\frac{1}{\Delta}\partial_i \partial_j\right) {h}^T+{h}_{i,j}+{h}_{j,i}, \end{equation}
where $1/\Delta$ is the inverse of the flat-space Laplacian.   Note that $h^T$ and $h_i$ can be found to be \cite{ADM}
\begin{equation} \begin{split}
h^T &= h_{ii} - \frac{1}{\Delta} h_{ij,ij} \\
h_i &= \frac{1}{\Delta} \left[ h_{ij,j} - \frac{1}{2\Delta}h_{kj,kji} \right].
\end{split} \end{equation}
With these new definitions in hand, we can write 
\begin{equation} \begin{split}
\int d^3x \pi^{ij}\partial_t g_{ij}  
&= \int d^3x \left[ (1 + {h}^T) \pi^{ii} \partial_t \log(1 + {h}^T) - 2 \pi^{ij}_{,j} \partial_t \left( {h}_i - \frac{1}{2\Delta} \partial_i {h}^T \right) \right] \\
&= \int d^3x \left[-\log(1 + {h}^T) \partial_t(\pi^{ii} (1 + {h}^T)) - 2 \pi^{ij}_{,j} \partial_t \left( {h}_i - \frac{1}{2\Delta} \partial_i {h}^T \right) \right] \\
&= \int d^3x \left\{ -\Delta \log(1+{h}^T) \partial_t \left[ \frac{1}{\Delta}((1+{h}^T)\pi^{ii})\right] - 2\pi^{ij}_{,j} \partial_t({h}_i - \frac{1}{2 \Delta} \partial_i {h}^T) \right\}
\end{split} \end{equation}
where we discarded total derivatives in arriving at the last expression.  Hence, the action (\ref{ac-1}), with constraints $R=0$ and $R^i=0$ from (\ref{eqn:split}), can be rewritten as
\begin{equation}
I = \int d^3x \sum_a p_{ai} \dot{z}_a^i \delta(\mathbf{x-z_a}) - \frac{1}{2 \kappa} \Delta \log(1 + {h}^T) \partial_0 \left[ \frac{1}{\Delta}((1+{h}^T)\pi^{ii})\right] - \frac{1}{\kappa} \pi^{ij}_{,j} \partial_0({h}_i - \frac{1}{2 \Delta} \partial_i {h}^T).
\end{equation}
This action suggests that we should adopt the coordinate conditions
\begin{equation} \label{cc-11}
	x^0 = -\frac{1}{\Delta} \left( \left[1+{h}^T\right]\pi^{ii} \right)  \hspace{1.cm}
	x^i ={h}_i-\frac{1}{2\Delta}\partial_i{h}^T,
\end{equation}
such that, defining the Hamiltonian density ${\cal H} = \frac{-1}{2 \kappa} \Delta \log(1+{h}^T)$ and using $x^\mu_{,\nu}=\delta^\mu_\nu$, the action can be rewritten as
\begin{equation}
	I=\int d^{3}x \left\{\sum_{a}p_{ai}\dot{z}^{i}_{a}\delta(\mathbf{x} -\mathbf{z}_{a})-{\cal H} \right\}.
\end{equation}
By imposing proper boundary conditions, we rewrite our coordinate conditions in a more convenient form:
\begin{equation} \begin{split}
\pi^{ii} = 0 \hspace{1.cm}
g_{ij,j} = \frac{1}{2}g_{jj,i}
\end{split} \end{equation}
where the second equation implies that $h_i - \frac{1}{2\Delta} \partial_i h^T=0$, which in turn means that $h_{ij} = \delta_{ij}h^T$, and thus, 
\begin{equation} g_{ij} = \delta_{ij}(1+h^T). 
\end{equation} 
In the coordinate system defined by these coordinate conditions and defining $\phi\equiv \mbox{log}(1+h^{T})$, the  constraint equations 
$R^{0}=0$ and $R^{i}=0$ lead to
\begin{gather}
\Delta\phi =-2\kappa\sum_a(m_a^2 +p^2_a e^{-\phi})^{\frac{1}{2}}\delta(\mathbf{x} -\mathbf{z}_a)-e^{\phi}\pi^{ij}\pi^{ij} \label{eq-phi}\\
\partial_j(\pi^{ij}e^{\phi}) = -\kappa\sum_ap_{ai}\delta(\mathbf{x} -\mathbf{z}_a), \label{eq-pi2}
\end{gather}
where the Hamiltonian is given by $H=-\frac{1}{2\kappa}\int d^2x \Delta \phi$ and where $p^2 \equiv \mathbf{p}^2$ is the square of the norm of the vector $\mathbf{p}$.  The solution satisfying the condition $\pi^{ii}=0$ is
\begin{equation}
	\pi^{ij}=e^{-\phi}\left\{-\frac{\kappa}{2\pi}\sum_{a}D_{ijk}(\mathbf{p_{a}}) \partial_{k}\;\mbox{log}\;r_{a}\right\},
\end{equation}
where $D_{ijk}(\mathbf{p_{a}})=p_{ai}\delta_{jk}+p_{aj}\delta_{ik}-p_{ak}\delta_{ij}$.  Thus, Equation (\ref{eq-phi}) can be rewritten as
\begin{eqnarray} \label{Dphi}
\Delta\phi&=&-2\kappa\sum_{a}\left(m_{a}^{2}+p_{a}^{2}
e^{-\phi(\mathbf{z_{a}})}\right)^{1/2}\delta(\mathbf{x}-\mathbf{z}_{a})
\nonumber \\
&&-\left\{\left(\frac{\kappa}{2\pi}\right)^{2}\sum_{a}\sum_{b}D_{ijk}(\mathbf{p_{a}})
D_{ijl}(\mathbf{p_{b}})\partial_{k}\;\mbox{log}\;r_{a}\partial_{l}\;\mbox{log}\;r_{b}\right\}\;e^{-\phi}.\label{eq-phi2}
\end{eqnarray}
We can then determine the Hamiltonian for a system of particles, which can be written down in the general form:
\begin{equation}\label{ham3}
H=\sum_{a}\left(m_{a}^{2}+p_{a}^{2}e^{-\phi(\mathbf{z_{a}})}\right)^{1/2}
+\frac{\kappa}{8\pi^{2}}\int d^{2}x \left\{\sum_{a}\sum_{b}D_{ijk}(\mathbf{p_{a}})
D_{ijl}(\mathbf{p_{b}})\partial_{k}\;\mbox{log}\;r_{a}\partial_{l}\;\mbox{log}\;r_{b}\right\}
e^{-\phi}\;\;.
\end{equation}
where  $\phi$ is obtained by solving (\ref{eq-phi2}). 

As will be exemplified in the following section, this Hamiltonian can be solved perturbatively in $\kappa$.  Writing $\phi = \sum \kappa^i \phi^{(i)}$, it is clear  that the first term of 
on the right-hand side of  (\ref{Dphi})
only depends on $\phi^{(1)}, \ldots, \phi^{(i-1)}$ while the second depends only on $\phi^{(1)}, \ldots, \phi^{(i-2)}$.  Thus, we inductively solve for each term in the power series for $\phi$ by calculating $\Delta \phi^{(i)}$ from the known $\phi^{(1)}, \ldots, \phi^{(i-1)}$, and solving it for $\phi^{(i)}$ before moving onto the $(i+1)$ term.  Provided that we can solve these $\Delta \phi$ differential equations, this method allows us to solve for the Hamiltonian up to any order in $\kappa$.

\section{Two-Particle System} \label{sec:1par}
To deal with the particles' divergent self-energies, which tend to be common in this sort of calculation, we introduce a renormalization scheme similar to that used in quantum field theory, consisting of introducing an energy scale to our measurement of distances.  We first introduce the density of particle $i$ to be 
\begin{equation}
	n_i(r) = \frac{a}{2 \pi r(r+a)^2}
\end{equation}
where $a$ is an intrinsic length scale depending on the particle's energy, and where we will be taking the limit as $a$ goes to zero.  Notice that for any value of $a$, $\int_{U} n dV = 1$ and $n(r)$ goes to a $\delta$ function as $a$ becomes small.  Moreover,  the equation $\Delta g_i = n_i(r)$ has the solution (up to a term with vanishing Laplacian)
\begin{equation}
	g_i = \frac{1}{2 \pi} \log \left( 1 + \frac{r_i}{a} \right) \equiv \frac{1}{2\pi}h(r_i),
\end{equation}
as opposed to simply $g \propto \log(r)$ were we to use the $\delta$ function.  Then, the self-energy of a particle will be $\log(1)=0$.  $a$ is to be thought of as the gravitational length scale of the interaction, which we assume to be on the order of the Planck length.  Since $a$ is dimensionful,  we can take the limit of it being a very small quantity, while at the same time choosing units such that $a=1$.  Then, for a non-zero $r$, $\log (1 + r/a)=\log(1+r)\approx \log(r)$.


We will consider a system of two particles of masses $m_1$ and $m_2$, in the centre of inertia frame where $\mathbf{p_1} = -\mathbf{p_2} = \mathbf{p}$, using the notation $E_a \equiv \sqrt{m_a^2+p_a^2}$ with, again, $p_a^2 \equiv \mathbf{p_a}^2$.  We also use shorthand notation $r_i = |\vecvar{x} - \vecvar{z_i}|$ and $r = |\vecvar{z_1} - \vecvar{z_2}|$.    

We first note that, in this frame,
\begin{equation}
	\sum_a D_{ijk}(\mathbf{p_a}) \partial_k h(r_a) = (p_i \partial_j + p_j \partial_i - \delta_{ij} p_k \partial_k)(h(r_1)-h(r_2))
\end{equation}																										
such that the Hamiltonian (\ref{ham3}) will have the form
\begin{equation}
	H = \sum_a \left( m_a^2 + p_a^2 e^{-\phi(\mathbf{z_a})} \right)^{1/2}
			+\frac{\kappa p^2}{4 \pi^2} \int d^2x \left[ \nabla (h(r_1)-h(r_2)) \cdot 
																										\nabla (h(r_1)-h(r_2)) \right] e^{-\phi}.
\end{equation}
We will be using a perturbative approach to third order in $\kappa$, and therefore calculate:
\begin{equation} \begin{split}
	\phi &= \kappa \phi^{(1)} + \kappa^2 \phi^{(2)} + \kappa^3 \phi^{(3)} \\
	e^{-\phi} &= 1- \kappa \phi^{(1)} - \kappa^2 \left( \phi^{(2)} - \frac{1}{2} \phi^{(1)} \phi^{(1)} \right)\\
\left( m_a^2+p_a^2 e^{-\phi} \right)^{1/2} &=  E_a - \frac{\kappa}{2} \frac{p_a^2}{E_a} \phi^{(1)} - \frac{\kappa^2}{2} \left( \frac{p_a^2}{E_a} \left( \phi^{(2)} - \frac{1}{2}\phi^{(1)}\phi^{(1)} \right) + \frac{(p_a^2)^2}{4E_a^3} \phi^{(1)} \phi^{(1)} \right).
\end{split} \end{equation}
Note that in the limit $\kappa \rightarrow 0$, we wish to retrieve $H = \sum_a \sqrt{m_a^2 + p_a^2}$, and therefore require $\phi$ to be of order at least $\kappa$. The contribution to the Hamiltonian will be separately calculated for each order in $\kappa$.  We find:
\begin{equation} \begin{split} \label{eqn:phi}
\Delta \phi^{(1)} &= -2 \left[ E_1 n_1(r_1) + E_2 n_2(r_2) \right] \\
\Delta \phi^{(2)} &= p^2 \left[ \frac{1}{E_1} \phi^{(1)}(\mathbf{z_1})n_1(r_1) + \frac{1}{E_2} \phi^{(1)}(\mathbf{z_2})n_2(r_2)\right] - \frac{p^2}{2\pi^2} \nabla (h(r_1)-h(r_2)) \cdot \nabla (h(r_1)-h(r_2)) \\
&= \frac{-p^2}{\pi} \left[ \left( \frac{E_2}{E_1}+1\right)h(r)n_1(r_1)
+ \left( \frac{E_1}{E_2}+1\right)h(r)n_2(r_2) \right]
 - \frac{p^2}{4\pi} \Delta \left[ h(r_1)-h(r_2) \right]^2 \\
\Delta \phi^{(3)} &= \frac{p^2}{E_1} \left[ \phi^{(2)}(\mathbf{z_1}) - \frac{1}{2} \phi^{(1)}(\mathbf{z_1})^2 + \frac{p^2}{4E_1^2} \phi^{(1)}(\mathbf{z_1})^2 \right] n_1(r_1) + (1 \leftrightarrow 2) + \frac{p^2}{2\pi^2} \left[ \nabla (h(r_1)-h(r_2)) \cdot \nabla (h(r_1)-h(r_2)) \right] \\
&= \frac{p^2}{E_1}\left[ \frac{-p^2}{2\pi^2} h^2(r) \left( 1 + \frac{E_1}{E_2}\right) - \frac{p^2}{4\pi^2}h(r)^2  - \frac{1}{2\pi^2}E_2^2 h(r)^2 + \frac{p^2}{4E_1^2\pi^2}E_2^2 h(r)^2 \right] n_1(r_1) \\
&~~~ + (1 \leftrightarrow 2) \\
&~~~ - \frac{p^2}{2\pi^3} \left[ \nabla (h(r_1)-h(r_2)) \cdot \nabla (h(r_1)-h(r_2)) \right] \left( E_1 h(r_1) + E_2 h(r_2) \right)
\end{split} \end{equation}
where we used the solutions
\begin{equation} \begin{split}
\phi^{(1)} &= \frac{-1}{\pi} \left[ E_1 h(r_1) + E_2 h(r_2) \right] \\
\phi^{(2)} &= \frac{-p^2}{2\pi^2} \left[ \left( \frac{E_2}{E_1}+1\right) h(r_1) h(r) + 
\left( \frac{E_1}{E_2}+1\right) h(r_2) h(r) \right] - \frac{p^2}{4\pi^2} \left( h(r_1)-h(r_2) \right)^2
\end{split} \end{equation}
The Hamiltonian $H = \frac{-1}{2} \int d^2x \Delta \phi$ can now be found as an expansion in $\kappa$:
\begin{equation} \begin{split}
H^{(0)} &= E_1 + E_2 \\
H^{(1)} &= \frac{p^2}{2\pi} \left[ \left( \frac{E_2}{E_1}+1 \right) \log(r) + 
\left( \frac{E_1}{E_2}+1 \right) \log(r) \right] \\
H^{(2)} &= \frac{p^2}{E_1} \log(r)^2 \left\{ \frac{p^2}{4\pi^2} \left( 1 + \frac{E_1}{E_2}\right) + \frac{p^2}{8\pi^2} + \frac{1}{4\pi^2} E_2^2  - \frac{p^2}{8E_1^2\pi^2}E_2^2 \right\} \\
&~~~ + (1 \leftrightarrow 2) \\
&~~~ + \frac{3p^2}{4\pi^2}(E_1 + E_2)\log(r)^2 .
\end{split} \end{equation}
where we used equations (\ref{app1}) and (\ref{app2}) and took the limit $h(r) \rightarrow \log(r)$.  In the case of equal masses, $E_1=E_2$, our Hamiltonian is greatly simplified:
\begin{equation} \label{HamEqualMasses}
H = 2E + \frac{2 \kappa p^2}{\pi} \log(r) + \frac{\kappa^2 p^2}{\pi^2 E} \log(r)^2 (p^2 + 2E^2)
\end{equation}
Simplifying further, the massless approximation gives:
\begin{equation} H = 2p + \frac{2\kappa}{\pi} p^2 \log r + \frac{3\kappa^2 p^3}{\pi^2} \log r ^2 \end{equation}
which we notice as being the second-order expansion of $\frac{-2\pi}{\kappa \log r} W \left( \frac{-\kappa}{\pi} p \log r\right)$ where $W$ is the Lambert $W$ function.  Note that the Hamiltonian in 1+1-dim dilaton gravity was also expressed in terms of the $W$ function \cite{Mann3}.

\section{Exact Quasi-Static Approximation}
\label{sec:quasistatic}
We can also use this formalism to describe exactly the Hamiltonian dynamics of a system in the quasi-static limit.  In the following, we will therefore retain all orders in the coupling constant $\kappa$ and only the lowest (linear) order in $p^{2}$.  A. Bellini et. al. treated the geodesic equations in this approximation \cite{Bellini}.  For simplicity we shall consider only the equal-mass case.  To simplify the notation, we will, in contrast to the previous section, return to a $\delta$ mass distribution and a $\log(r)$ potential.  We will explicitly keep the divergent terms $\log r_{11}$ and $\log r_{22}$, where $r_{ii}=|x_i-x_i|$, to display their cancellation at every order in $\kappa$.  From (\ref{ham3}) the Hamiltonian in this approximation is given by
\begin{equation} \begin{split} \label{ham6}
	H =&2m + \frac{p^2}{2m} e^{\frac{\kappa}{\pi}m \log r_{12}} \left( e^{ \frac{\kappa}{\pi}m \log r_{11}} + e^{ \frac{\kappa}{\pi}m \log r_{22}} \right) \\
	&~~	+ \frac{\kappa p^2}{4\pi^2}\int d^2x \left(	\nabla(\log r_1-\log r_2)\cdot \nabla (\log r_1-\log r_2)\right) e^{\frac{\kappa}{\pi}m(\log r_1 +\log r_2)}.
\end{split} \end{equation}
where $\phi_s=-\frac{\kappa}{\pi}\sum_a \log r_a$ is a static solution to (\ref{eq-phi2}) and where $m_{1}=m_{2}=m$ and $\vecvar{p}_1=-\vecvar{p}_2=\vecvar{p}$.  The integral of the 3rd term on the right hand side of (\ref{ham6}) is
\begin{equation} \begin{split}
\int d^2x & \left[ \nabla(\log r_1 - \log r_2) \cdot \nabla(\log r_1 - \log r_2) \right] e^{\frac{\kappa}{\pi} m (\log r_1 + \log r_2)} \\
&= \int d^2x \left[ \nabla(\log r_1 + \log r_2) \cdot \nabla(\log r_1 + \log r_2) - 4 (\nabla \log r_1 \cdot \nabla \log r_2) \right] e^{\frac{\kappa}{\pi} m (\log r_1 + \log r_2)}\\
&= \frac{-2 \pi^2}{\kappa m} \left[ e^{\frac{\kappa}{\pi}m (\log r_{11}+\log r_{12})} +e^{\frac{\kappa}{\pi}m (\log r_{12}+\log r_{22})}  - 2 \right] \\
&~~~ - 4 \int d^2x (\nabla \log r_1 \cdot \nabla \log r_2) e^{\frac{\kappa}{\pi} m (\log r_1 + \log r_2)}.
\end{split} \end{equation}
The first term in the preceding expression cancels the second (divergent) term in eq. (\ref{ham6}).
Using the integration formula
\begin{equation} \int d^2x (\nabla \log r_1 \cdot \nabla \log r_2)(\log r_1 + \log r_2)^n = -\frac{2^{n+1}}{n+1}\pi (\log r)^{n+1}, \end{equation}
we can rewrite the Hamiltonian (\ref{ham6}) as
\begin{equation} \begin{split}
H &= 2m + \frac{p^2}{m} - \frac{\kappa p^2}{\pi^2} \int d^2x (\nabla \log r_1 \cdot \nabla \log r_2) e^{\frac{\kappa}{\pi} m (\log r_1 + \log r_2)} \\
&= 2m + \frac{p^2}{m} e^{\frac{2 \kappa m}{\pi} \log r} \\
&= 2m + \frac{p^2}{m} (r^2)^{\frac{\kappa m}{\pi}}
\end{split} \end{equation}
which indeed agrees with equation (\ref{HamEqualMasses}) in the quasi-static limit.  This Hamiltonian gives us the canonical equations of motion
\begin{eqnarray}
\dot{\vecvar{r}}&=&\frac{\partial H}{\partial \vecvar{p}}
=\frac{2}{m}(r^{2})^{\frac{\kappa m}{\pi}}\;\vecvar{p}
\\
\dot{\vecvar{p}}&=&-\frac{\partial H}{\partial \vecvar{r}}
=-\frac{2\kappa}{\pi} p^{2}(r^{2})^{\frac{\kappa m}{\pi}-1}\;\vecvar{r},
\end{eqnarray}
which lead to the equations of motion in the second order
\begin{equation}
\ddot{\vecvar{r}}=\frac{\kappa m}{\pi}\frac{2\dot{\vecvar{r}}(\vecvar{r}\cdot
\dot{\vecvar{r}})-\vecvar{r}(\dot{\vecvar{r}})^{2}}{\vecvar{r}^{2}}\;\;.
\end{equation}
This equation is identical with Eq.(4.7) in \cite{Bellini}: $\ddot{\eta}_{r}=4GM\;\frac{(\dot{\eta}_{r})^{2}}{\eta_{r}}$, where $\eta_{r}=\eta_{x}+i\;\eta_{y}, G=\frac{\kappa}{8\pi}$ and $M=2m$.

\section{Conclusion}

We have presented here the general expression (\ref{ham3}) for the Hamiltonian describing $N$ particles coupled to (2+1) dimensional gravity.   Beginning with a derivation a general form for the action corresponding to $N$-body dynamics, we employed the ADM formalism and proper coordinate conditions  to find this general Hamiltonian.   Our results are complementary to those that employ topological methods \cite{Bellini,Bellini2,Matschull}.   

Our result  (\ref{ham3}) has the advantage that it can be studied in a wide variety of physical regimes, including small and large mass, small and large momenta, and small and large gravitational coupling.
We explicitly demonstrated this in two cases.  First we obtained
an expansion to second-order in the gravitational coupling constant $\kappa$ for the two-particle system. Second we also considered  a quasi-static, two-particle equal-mass system to every order in $\kappa$, in which case our results agree with those previously found in the literature \cite{Bellini,Bellini2}.

A number of interesting problems remain for future consideration.  An obvious thing to try is to incorporate additional couplings to electromagnetism and a cosmological constant.  A study of the quantization of the Hamiltonian (\ref{ham3}) should also afford interesting insight into the nature of (2+1) quantum gravity coupled to matter.

\section*{Acknowledgements}

This work was supported  by the Natural Sciences and Engineering Council of Canada.

\section{Appendix: Fourier Integrals}
This appendix will contain proofs of some integration formulas used in section \ref{sec:1par}, which are done through Fourier transforms.  We begin by noticing that
\begin{equation} \frac{-1}{2\pi}\int d^2k \Delta \frac{1}{k^2} e^{ik \cdot x} = \delta(x) \end{equation}
and therefore, the two-dimensional Fourier transform of $\log(r)$ is $\frac{-2\pi}{k^2}$.  We use this to show that, for $n \geq 2$, 
\begin{equation} \begin{split} \label{app1}
\int d^2x \Delta (\log r_1 \ldots \log r_n) &= \frac{1}{(2\pi)^n} \int d^2x \Delta \int \prod_{i=1}^n \frac{d^2k_i}{k_i^2}e^{i\sum k_i \cdot x} e^{-i \sum_i k_i \cdot z_i} \\
&= \frac{1}{(2\pi)^{n-2}} \int \prod_{i=1}^n \frac{d^2k_i}{k_i^2} \left(\sum k_i \right)^2 \delta \left(\sum k_i \right) e^{-i \sum_i k_i \cdot z_i} \\
&= 0,
\end{split} \end{equation}
meaning that the integral of the laplacian of any polynomial function in $\log r$ will vanish.  Similarly, we can show
\begin{eqnarray} \label{app2}
\lefteqn{\int d^{2}x ({\bf\nabla}\;\mbox{log}\;r_{1}\cdot{\bf\nabla}\;\mbox{log}\;r_{2})\;\mbox{log}\;r_{3}}
\nonumber \\
&=&\frac{1}{(2\pi)^{3}}\int d^{2}xd^{2}k_{1}d^{2}k_{2}d^{2}k_{3}\;
\frac{(\vecvar{k}_{1}\cdot\vecvar{k}_{2})}{\vecvar{k}^{2}_{1}\vecvar{k}^{2}_{2}\vecvar{k}^{2}_{3}}\;e^{i(\vecvar{k}_{1}+\vecvar{k}_{2}+\vecvar{k}_{3})\cdot
\vecvar{x}}e^{-i(\vecvar{k}_{1}\cdot\vecvar{z}_{1}+\vecvar{k}_{2}\cdot\vecvar{z}_{2}+\vecvar{k}_{3}\cdot\vecvar{z}_{3})}
\nonumber \\
&=&\frac{1}{2\pi}\int d^{2}k_{1}d^{2}k_{2}d^{2}k_{3}\;
\frac{(\vecvar{k}_{1}\cdot\vecvar{k}_{2})}{\vecvar{k}^{2}_{1}\vecvar{k}^{2}_{2}\vecvar{k}^{2}_{3}}\;\delta(\vecvar{k}_{1}+\vecvar{k}_{2}+\vecvar{k}_{3})
\;e^{-i(\vecvar{k}_{1}\cdot\vecvar{z}_{1}+\vecvar{k}_{2}\cdot\vecvar{z}_{2}+\vecvar{k}_{3}\cdot\vecvar{z}_{3})}
\nonumber \\
&=&\frac{1}{4\pi}\int d^{2}k_{1}d^{2}k_{2}d^{2}k_{3}\;
\left(\frac{1}{\vecvar{k}^{2}_{1}\vecvar{k}^{2}_{2}}
-\frac{1}{\vecvar{k}^{2}_{2}\vecvar{k}^{2}_{3}}
-\frac{1}{\vecvar{k}^{2}_{1}\vecvar{k}^{2}_{3}}\right)
\delta(\vecvar{k}_{1}+\vecvar{k}_{2}+\vecvar{k}_{3})\;
e^{-i(\vecvar{k}_{1}\cdot\vecvar{z}_{1}+\vecvar{k}_{2}\cdot\vecvar{z}_{2}+\vecvar{k}_{3}\cdot\vecvar{z}_{3})}
\nonumber \\
&=&\pi\left(\;\mbox{log}\;r_{31}\;\mbox{log}\;r_{32}
-\mbox{log}\;r_{12}\;\mbox{log}\;r_{13}
-\mbox{log}\;r_{21}\;\mbox{log}\;r_{23}\right)\;\;.
\nonumber 
\end{eqnarray}
which implies that
\begin{equation} \begin{split}
\int d^2x (\nabla \log r_1 \cdot \nabla \log r_2) \log r_1 &= -\pi (\log r)^2 \\
\int d^2x (\nabla \log r_1 \cdot \nabla \log r_2) \log r_2 &= -\pi (\log r)^2 \\
\end{split} \end{equation}
These integrals will not change when we make the replacement $\log(r) \rightarrow \log(1 + r/a)$ for suitably small $a$, as they get little contribution for the small $r$ part of the integral.  This can be seen by noticing that replacing the bounds of integration by $\int _0^\epsilon$ makes the integrals vanish.

\end{document}